# New energy definition for higher-curvature gravities

S. Deser[*]

*California Institute of Technology, Pasadena, California 91125 USA and Brandeis University, Waltham, Massachusetts 02454, USA*

Bayram Tekin[†]

*Department of Physics, Faculty of Arts and Sciences, Middle East Technical University, 06531, Ankara, Turkey*

We propose a novel but natural definition of conserved quantities for gravity models of quadratic and higher order in curvature. Based on the spatial asymptotics of curvature rather than of metric, it avoids the more egregious problems—such as zero-energy "theorems" and failure in flat backgrounds—in this fourth-derivative realm. In $D > 4$, the present expression indeed correctly discriminates between second-derivative Gauss-Bonnet and generic, fourth-derivative actions.

## I. INTRODUCTION

The definitions [1,2] of conserved charges in general relativity (GR)—without or with a cosmological term—are both simple and natural for the study of asymptotically flat or constant curvature solutions.

Being a gauge theory, GR has constraint equations and Bianchi identities. Together with the Killing vectors associated to the asymptotic geometry at spatial infinity, they provide an intuitively simple physical framework for defining the generators of Poincare/de Sitter (dS) transformations, generalizing the Poisson equation in Abelian electrostatics. They also identify the values of those conserved quantities characterizing isolated systems in terms of the spatial asymptotics of (invariant components of) the metric.

Fitting quadratic curvature models into the same mold proves difficult because (being of fourth-derivative order, in general) these typically have constraints $\sim \nabla^4 \phi = \rho$ with (naively at least) rising, $\phi \sim Q r^{+1}$, solutions. While some gymnastics (for which the present authors are responsible [3,4]) can accommodate defining conserved quantities in a parallel way to GR, they do so at a considerable cost. First, one must use asymptotically de Sitter (rather than flat) spaces to obtain any nonzero $E$ at all; second, some perfectly normal quadratic models acquire zero energy. A related paradox in this direction is the (in)famous zero-energy "theorem" [5] of conformal (Weyl) gravity in $D = 4$. That one is based on the fact that the coefficient of the metric's $1/r$ (in $D = 4$) is no longer proportional to energy as in GR. The formal solution provided in [3,4] is also unsatisfactory: physically what is needed, and provided here, is the "Poisson equation" whose (leading term) source is the matter stress tensor.

The above problems disappear in the present, simple framework; there will be neither "zero-energy" models nor a need to appeal to dS versus flat backgrounds. For simplicity, we will mainly work in $D = 4$ and asymptotically flat background, but will also indicate how to include dS asymptotics, and the special role of Gauss-Bonnet (GB) gravitation in $D > 4$ will confirm our definition.

## II. THE MODELS

For present purposes, the relevant theories are of the form

$$16\pi\kappa I = \int d^4 x \sqrt{-g} (\alpha R^2 + \beta R_{\mu\nu}^2), \quad (1)$$

with conventions $(-+++)$, $R_{\mu\nu} \equiv R^\alpha_{\mu\alpha\nu} \sim -1/2 \Box h_{\mu\nu}$, in the expansion

$$g_{\mu\nu} \equiv \bar{g}_{\mu\nu} + h_{\mu\nu} \quad (2)$$

about backgrounds $\bar{g}_{\mu\nu}$. In (and only in) $D = 4$, the third possible quadratic term, $\gamma \int d^4 R_{\mu\nu\alpha\beta}^2$, is irrelevant owing to the GB identity:

$$\delta \int d^4 x [R_{\mu\nu\alpha\beta}^2 - 4 R_{\mu\nu}^2 + R^2] = 0.$$

In fact, we will see that this term fits nicely into the story beyond $D = 4$; likewise, higher powers of curvature do not contribute to the left-hand side (LHS) of the constraint equations about flat $\bar{g}_{\mu\nu}$. The field equations that follow from (1) and all other relevant details may be found in [3,4], whose notation we follow here. In particular, we reproduce these equations and their expansion about a constant curvature background. The field equation is

$$2\alpha R(R_{\mu\nu} - \tfrac{1}{4} g_{\mu\nu} R) + (2\alpha + \beta)(g_{\mu\nu}\Box - \nabla_\mu \nabla_\nu) R$$
$$+ \beta \Box (R_{\mu\nu} - \tfrac{1}{2} g_{\mu\nu} R) + 2\beta (R_{\mu\sigma\nu\rho} - \tfrac{1}{4} g_{\mu\nu} R_{\sigma\rho}) R^{\sigma\rho} = \kappa \tau_{\mu\nu}, \quad (3)$$

where $\tau_{\mu\nu}$ is the matter source and the LHS of (3) obeys a "Bianchi" identity, being the variation of a diffeo-invariant action. If we now cast all terms nonlinear in $h_{\mu\nu}$ into the

---

[*]Electronic address: deser@brandeis.edu
[†]Electronic address: btekin@metu.edu.tr

"source" side of the equations, then (3) becomes

$$(2\alpha + \beta)(\bar{g}_{\mu\nu}\Box - \bar{\nabla}_\mu \bar{\nabla}_\nu + \Lambda g_{\mu\nu})R_L$$
$$+ 4\Lambda\left(2\alpha + \frac{\beta}{3}\right)G^L_{\mu\nu} + \beta\Box G^L_{\mu\nu} - \frac{2\beta\Lambda}{3}\bar{g}_{\mu\nu}R_L = T_{\mu\nu}, \quad (4)$$

where all operations are with respect to the background $\bar{g}_{\mu\nu}$, and the linearized Einstein tensor obeys $\bar{\nabla}^\mu G^L_{\mu\nu} \equiv 0$. (The $T_{\mu\nu}$ on the RHS of (4) is the sum of matter and nonlinear gravitational contributions.) The field equations (3) admit source-free solutions, $\bar{g}_{\mu\nu}$ of constant curvature $\bar{R}_{\mu\sigma\nu\rho} = \Lambda/3(\bar{g}_{\mu\nu}\bar{g}_{\sigma\rho} - \bar{g}_{\nu\rho}\bar{g}_{\mu\sigma})$, $\bar{R}_{\mu\nu} = \Lambda\bar{g}_{\mu\nu}$, for any $\Lambda$, so the definitions we are seeking must be valid for all dS spaces; however, flat background is most transparent for explaining their properties; the extension to nonzero $\Lambda$ is straightforward. Our (4) simplifies, at $\Lambda = 0$, to

$$(2\alpha + \beta)(\bar{g}_{\mu\nu}\Box - \bar{\nabla}_\mu \bar{\nabla}_\nu)R_L + \beta\Box G^L_{\mu\nu} = T_{\mu\nu}. \quad (5)$$

Henceforth we will drop the "$L$" index on the linear curvatures.

## III. ENERGY DEFINITION

Consider just the (00) energy component of (5). For GR, the LHS is just $G_{00} \equiv -\frac{1}{2}\nabla^2 h^T$ where $h^T$ is the "transverse-trace" component of $h_{ij}$ in the usual orthogonal decomposition. There, one immediately notes that $h^T \sim \int T_{00} d^3x/r = E/r$ where $E$ is guaranteed to be conserved (because $\partial_0 G_{00} \equiv \partial_i G_{0i}$), and similarly for the (0$i$) and angular momentum constraints. Just as the Einstein constraint $G_{00}$ has no time derivatives, compared to the two derivatives in the time evolution equations, here we would expect the (00) component of (5) to have two, rather than the four, derivatives in its evolution equations. This is indeed the case, because the projector $P_{\mu\nu} \equiv (\bar{g}_{\mu\nu}\Box - \bar{\nabla}_\mu \bar{\nabla}_\nu)$ obeys $P_{00} \equiv -\nabla^2$. Hence the energy constraint becomes

$$-\nabla^2[(2\alpha + \beta)R + \beta/2\Box h^T] = T_{00}. \quad (6)$$

[Both $R$ and $\Box h^T$ indeed have (noncanceling) $\partial_0^2$ terms, as expected, which poses no obstacle, since they are simply incorporated into the defining linearized curvatures. We could even rewrite the $h^T$ term more "invariantly" as

$$\Box h^T \equiv (\nabla^2 - \partial_0^2)h^T = -2G_{00} + 2(\partial^2_{ij}\nabla^{-2})G_{ij},$$

using the Bianchi identity $\bar{\nabla}_\mu \bar{\nabla}_\nu G^{\mu\nu} \equiv 0$.] Hence we conclude from (6) that the curvature combination in the square brackets is a "Poisson potential," behaving as $(\int d^3 r T_{00})/r \equiv E/r$ at spatial infinity: Unsurprisingly, in these fourth-derivative models, second derivatives of $h_{\mu\nu}$ act as potentials (rather than $h^T$ itself in second order GR). This is our basic result,

$$[(2\alpha + \beta)R + \beta/2\Box h^T] \sim \int d^3 r T_{00}/r \equiv E/r. \quad (7)$$

We immediately note that, unlike the "traditional" definition of [3,4], this one is already valid for flat background and $E$ does not vanish for any $(\alpha, \beta)$ combination including the old [3,4], "bad" one $4\alpha + \beta = 0$, and Weyl gravity's $\beta + 3\alpha = 0$ [5]. We also note that the solutions here have very "weak" asymptotics—the curvatures vanish as slowly as $1/r$ at spatial infinity, which precludes $1/r$ behavior of $h_{\mu\nu}$ (but need not imply that $h_{\mu\nu} \sim r^{+1}$, since space is still flat out there.) We will not analyze the momentum constraint, since the message of (7) is clear—the constants of motion are carried by the leading asymptotic terms in (appropriate components of) the curvatures. That the charges are conserved follows from the linearized Bianchi identities obeyed by (5), using the (here "invisible") asymptotic Killing vectors $\bar{\xi}_\mu$: $\bar{\nabla}_\mu \bar{\xi}_\nu + \bar{\nabla}_\nu \bar{\xi}_\mu = 0$ to convert a tensor (density) conservation law $\bar{\nabla}_\mu G^{\mu\nu} = 0$ to the vector one $\partial_\nu(\bar{\xi}_\mu G^{\mu\nu}) \equiv \partial_\nu J^\nu = 0$, thereby providing an ordinarily conserved vector current. Exactly the same procedure leads to the conserved charges in asymptotically dS backgrounds, simply by keeping all terms in (4) and using the Killing vectors of dS.

We now clarify some important points. First, possible higher power additions, $\Delta I \sim \int d^D x R^n \quad n > 2$, do not change things: In flat background, they only contribute to the source side of the constraints because their linearizations vanish there. In constant curvature backgrounds, where $\bar{R}_{\mu\nu\alpha\beta}$ is nonzero, there will be linear terms of the form $\sim \Lambda^{(n-2)}\bar{\nabla}\bar{\nabla} R$, which only alters the details. Turning to $D > 4$, the main difference is the possible presence of the now independent quadratic term $\gamma \int d^D x R^2_{\mu\nu\alpha\beta}$. The latter will vary into two parts, $\sim (\nabla\nabla R)_{\mu\nu} + (RR)_{\mu\nu}$ (just as the $R^2_{\mu\nu}$ and $R^2$ actions did). In flat backgrounds, the $(RR)$ term becomes part of the $T_{\mu\nu}$, while $(\nabla\nabla R)$ is part of the "Poisson" LHS. Without even calculating, it is clear that, since $(\nabla\nabla R)_{\mu\nu}$ involves two contractions between the derivative and curvature indices, it must be a linear combination of the two terms in (5), these being the only identically conserved tensors of this order. Indeed, a simple calculation shows that

$$\gamma\delta\int d^D x R^2_{\mu\nu\alpha\beta} \equiv 2\gamma(\bar{\Box}G_{\mu\nu} + P_{\mu\nu}R). \quad (8)$$

This in turn tells us that the only chance of "zero energy" here is that of the GB action $\sim \int d^D x (R^2_{\mu\nu\alpha\beta} - 4R^2_{\mu\nu} + R^2)$, which is gratifying; among all $R^2$ models, only the GB field equations are of second-derivative order, so it is the one case to which our definition should not apply. Indeed, this system has asymptotically Schwarzschild-dS solutions [6], whose energy is quite adequately expressed by the usual GR definitions [3,4]. We emphasize that there necessarily had to be one (and only one) quadratic curvature model with vanishing "curvature" energy, since there are three parameters ($\alpha$, $\beta$, $\gamma$) (of which one is an overall scale) and only two independent tensors in (5) for general $D$. That GB has a special role in this framework provides an additional argument in favor of the curvature definition in the generic case.

Finally, we consider the GR + $R^2$ models, in which the Einstein term of pure GR is also present in (1). Aside from its nonlinear contributions to $T_{\mu\nu}$, it adds the normal $\kappa^{-2}G_{\mu\nu}$ to the LHS of (4), and hence the term $\kappa^{-2}G_{00} = -\kappa^{-2}/2\nabla^2 h^T$ to (5), giving rise to the combination $(\beta\Box + \nabla^2)h^T$ in (6). Formally this means that (7) acquires the usual metric contribution [1], $\kappa^{-2}h^T$, rather than being entirely curvature dependent. So, for solutions of the usual asymptotic Schwarzschild form, $h^T \sim m/r$, we may fall back on the GR definition of $E$, since the falloff is once again $h_{\mu\nu} \sim (1/r)$.

## IV. SUMMARY

We have proposed a new definition of conserved quantities in asymptotically flat or dS solutions of quadratic (and higher) curvature actions in any $D$. It is both simpler and better adapted to these models than attempting to follow the definitions valid for GR and removes the latter's paradoxes, including the zero-energy disease and failure at flat background. At the same time it faithfully adheres to the physics of the GR paradigm.

## ACKNOWLEDGMENTS

S. D. was supported in part by NSF Grant No. PHY04-01667. B. T. is partially supported by the Turkish Academy of Sciences (TÜBA) and by the TÜBİTAK Kariyer Grant No. 104T177.